\documentclass[twocolumn,aps,prl,showpacs]{revtex4}
\usepackage[dvips]{graphicx}
\begin{document}
\title{Pairbreaking Without Magnetic Impurities in Disordered Superconductors}
\author{Amit K. Chattopadhyay}
\email{akc@mpipks-dresden.mpg.de}
\author{Richard A. Klemm}
\email{rklemm@mpipks-dresden.mpg.de}
\author{Debanand Sa}
\email{deba@mpipks-dresden.mpg.de}
\affiliation{Max-Planck-Institut f{\"u}r Physik komplexer Systeme,
N{\"o}thnitzer
 Stra{\ss}e 38, D-01187 Dresden, Germany}
\date{\today}

\begin{abstract}
We study analytically the effects of inhomogeneous   pairing
interactions in  short coherence length superconductors, using a
spatially varying 
Bogoliubov-deGennes model.   Within the Born
approximation, it reproduces all of the standard
Abrikosov-Gor'kov pairbreaking and gaplessness effects,  even in the absence
of actual magnetic impurities.     
For  pairing disorder on a single site, the T-matrix  gives rise to bound  states within the BCS
gap.   Our results are compared with recent
scanning tunneling microscopy measurements on
Bi$_2$Sr$_2$CaCu$_2$O$_{8+\delta}$ with Zn or Ni impurities.
\end{abstract}
\vskip0pt\vskip0pt
\maketitle  
The effect of disorder in superconductors has long been a subject of 
considerable interest. 
A generally accepted physical picture is that magnetic impurities 
destroy superconductivity by locally breaking the pairs,\cite{AG,deg64,amb65,pa69}  
whereas non-magnetic impurities are not pairbreaking,  
according to Anderson's theorem.\cite{pa59}
This is true for an isotropic s-wave 
BCS superconductor, in which the order parameter is uniform and momentum independent. 
  Since in most high transition temperature $T_c$ cuprates, the
suppression of $T_c$ with  Zn or  Ni doping is comparable, \cite{mah94} there were
proposals to explain this in terms of  $d$-wave
superconductivity.\cite{sal96} 
However, nuclear magnetic resonance 
(NMR) experiments indicated that the
nominally non-magnetic Zn$^{++}$ ions polarize the spin background in the
CuO$_2$ planes upon substitution of the nominally $S=1/2$ Cu$^{++}$ sites.\cite{mah94,julien00} Moreover, recent scanning tunneling microscopy (STM) 
measurements directly above the Zn or Ni impurity sites observed strong
resonance peaks.\cite{pan00,hudson01}  Very
recently, several groups  noticed from STM measurements that  underdoped
Bi$_2$Sr$_2$Cu$_2$O$_{8+\delta}$ (BSCCO) and Bi$_{2-x}$Pb$_x$Sr$_2$Cu$_2$O$_{8+\delta}$ are extremely disordered on a scale of a few nm. \cite{pan01,lang02,HFK,Cren}
This disorder is characterized by two gaps, one corresponding to the
superconducting gap, with characteristic superconducting peaks, and a
non-superconducting gap.

There is now a large body of evidence that the pseudogap observed in cuprates above $T_c$ is not superconducting. The most convincing of these
experiments determined that the magnetic field dependence of the resistivity,
NMR spin-lattice relaxation rate, magnetization, and the gaps observed in
intrinsic tunneling experiments for the superconducting and pseudogaps are
qualitatively different.\cite{Shibauchi,SW,lang02,Krasnov,Zheng,Klemm}
In particular, the pseudogap regime is field independent until one reaches the Zeeman field for breaking up
chargeless spin zero pairs.\cite{Shibauchi}  This is precisely consistent with
the pseudogap being particle-hole pairs, which could be either of the
charge-density wave (CDW) or spin-density wave (SDW) form.  As in one
dimension, such excitations are expected to arise from repulsive interactions
between like charges, whereas superconductivity can arise when the
interactions between like charges are attractive.  

Thus, if indeed the disorder involves a percolation problem between
superconducting and density-wave regions on a nanometer scale, then the phase
coherence can only arise from Josephson coupling of the superconducting
grains.  In addition, one would expect the $c$-axis tunneling to be very
incoherent, as inferred in BSCCO from $c$-axis twist Josephson junction experiments. \cite{Li} 

In this letter, we assume the superconductor is electronically disordered
on the scale of the coherence length.  We
further assume the essential ingredient in the disorder is not one of
impurities, but rather disorder in the pairing interaction itself.  Thus, we
expect that $T_c$ varies from site to site, as does the
resulting order parameter amplitude.\cite{OWK}  We  treat this type of 
disorder
 using a Bogoliubov-deGennes procedure, assuming that the order parameter
amplitude varies locally. \cite{HS,Nandini}

Here we show that in the Born approximation, this problem   
maps exactly onto that of  pairbreaking in a superconductor,  with all of
the  features of that model, including 
gapless superconductivity. \cite{AG,deg64,pa69,amb65} At a particular defect site, the T-matrix gives
rise to  bound states within the gap, even without magnetic impurities.

We use the Nambu representation,
\begin{eqnarray}
\Psi({\bf r})&=&\pmatrix{c_{\uparrow}({\bf r})\cr
c_{\downarrow}({\bf r})\cr
c_{\uparrow}^{\dag}({\bf r})\cr
c_{\downarrow}^{\dag}({\bf r})},\nonumber\\
\Psi^{\dag}({\bf r})&=&\pmatrix{c_{\uparrow}^{\dag}({\bf
r}),&c_{\downarrow}^{\dag}({\bf r}),&c_{\uparrow}({\bf r}),&c_{\downarrow}({\bf r})},
\end{eqnarray} 
where $c_{\downarrow}({\bf r})$ [$c_{\uparrow}^{\dag}({\bf r})$] annihilates
[creates] a quasiparticle with spin eigenstate $\downarrow$ [$\uparrow$] at the
position ${\bf r}$.  We set $\hbar=c=k_B=1$.
The Hamiltonian under study is $H=H_0+H_1+H_2+H_3$, where
\begin{eqnarray}
H_0&=&\int d^3{\bf r}\Psi^{\dag}({\bf r})[\hat{\xi}({\bf
r})\rho_3\sigma_0+\Delta_0\rho_2\sigma_2]\Psi({\bf r}),\\
H_i&=&{1\over{2}}\int d^3{\bf r}\Psi^{\dag}({\bf r})\hat{V}_i({\bf r})\Psi({\bf
r}),\\
\hat{V}_1({\bf r})&=&U_1({\bf r})\rho_3\sigma_0,\\
\hat{V}_2({\bf r})&=&U_2({\bf r}){\bf
S}\cdot\overrightarrow{\alpha}/[S(S+1)]^{1/2}\nonumber\\
\overrightarrow{\alpha}&=&\hat{\bf x}\rho_3\sigma_1+\hat{\bf
y}\rho_0\sigma_2+\hat{\bf z}\rho_3\sigma_3,\\
\hat{V}_{3}({\bf r})&=&U_3({\bf r})\rho_2\sigma_2,
\end{eqnarray}
where $\rho_j\sigma_{j'}\equiv\rho_j\otimes\sigma_{j'}$  is a rank 4 tensor composed of two 
Pauli matrices for $j,j'=1,2,3$ and $\rho_0$, $\sigma_0$ are rank 2 identity
matrices, respectively.  $H_0$ is the Bogoliubov-deGennes
version of the BCS Hamiltonian, with momentum space quasiparticle energy
dispersion $\xi_{\bf k}$ relative to the Fermi energy $\mu$, $\Delta_0(T)$ is
the real bare uniform BCS order parameter, and $H_1$ and $H_2$
are the interactions due to scattering off random non-magnetic and magnetic
impurities  with  effective  potentials $U_1({\bf r})$ and
$U_2({\bf r})/[S(S+1)]^{1/2}$ , respectively.  $H_{3}$ with effective
potential $U_3({\bf r})$ is the new interaction arising from
random variations in the pairing interaction.\cite{OWK,Nandini}
In $H_2$, ${\bf S}$ and $S$ are the  spin vector and quantum number of 
the magnetic
impurities, respectively, and
$\overrightarrow{\alpha}$ 
represents the quasiparticle spin eigenvector.  We assume the spatial
averages of all of the random potentials vanish, $\langle
U_i({\bf r})\rangle=0$ for $i=1, 2, 3$.   In the absence of all defects,
the order parameter $\Delta_0(T)=V\langle c_{\uparrow}({\bf
r})c_{\downarrow}({\bf r})\rangle$ satisfies the standard BCS gap equation,
\begin{eqnarray}
\Delta_0&=&-VT\sum_{|\omega_n|\le\omega_0}\int {{d^3{\bf
k}}\over{(2\pi)^3}}{\rm Tr}[\rho_2\sigma_2\hat{G}_0({\bf k},\omega_n)],\nonumber\\
\hat{G}_0^{-1}({\bf k},\omega_n)&=&i\omega_n\rho_0\sigma_0-
\xi_{\bf k}\rho_3\sigma_0-\Delta_0\rho_2\sigma_2,
\end{eqnarray}
where $\hat{G}_0$ is the bare Green's function matrix, $V<0$ is the  uniform (BCS) part of the  pairing interaction, $N(0)$ is 
the single-spin
quasiparticle density of states, $\omega_0$ is a BCS-like cutoff, and $\omega_n$ are the Matsubara frequencies.

We   assumed a real bare uniform order parameter $\Delta_0$, and
restricted our consideration  in $H_3$ to spatial fluctuations of the
amplitude of $\Delta_0$.  The model can  also treat spatial
fluctuations of the  phase of $\Delta_0$ by letting $\Delta_0\rho_1\sigma_2$
and $U_3({\bf r})\rho_2\sigma_2$ be generalized to
$\Delta_{01}\rho_1\sigma_2+\Delta_{02}\rho_2\sigma_2$ and $U_{31}({\bf
r})\rho_1\sigma_2+U_{32}({\bf r})\rho_2\sigma_2$, respectively.

Our main interest lies in
studying $H_3$. Using quantum Monte Carlo techniques to study a
two-dimensional square lattice with an on-site attractive Hubbard pairing 
interaction
in $H_0$,\cite{Nandini} Ghosal {\it et al.} obtained interesting results in 
excellent qualitative agreement with those obtained from  STM
measurements.\cite{lang02}  We also consider $H_1$ and $H_2$ for comparison,
 because the combination of one or both of them with $H_3$ can lead to
interesting novel behavior.  In the Born
approximation, these interactions add or subtract in a simple
fashion.  However, in the T-matrix approximation for a single defect site,
these interactions interact in a highly non-trivial manner.

In the self-consistent Born approximation,
the quasiparticle self-energy $\hat{\Sigma}=\hat{\Sigma}_1+\hat{\Sigma}_2+\hat{\Sigma}_3$, where
\begin{eqnarray}
\hat{\Sigma}_i({\bf k},\omega_n)&=&n_i\sum_{{\bf k}'}\hat{V}_i({\bf k}-{\bf
k}')\hat{G}({\bf k}',\omega_n)\hat{V}_i({\bf k}'-{\bf k}),\>\>\\
\hat{G}^{-1}({\bf k},\omega_n)&=&\hat{G}^{-1}_0({\bf
k},\omega_n)-\hat{\Sigma}({\bf k},\omega_n),
\end{eqnarray}
 $\hat{V}_i({\bf k}), U_i({\bf k})$ are the
spatial Fourier transforms of $\hat{V}_i({\bf r}), U_i({\bf r})$, 
respectively.  Neglecting any possible anisotropy arising
from Fermi surface integrations, the effective rates of
the three processes are
\begin{eqnarray}
1/\tau_i&=&2\pi n_iN(0)|U_i({\bf k}_F)|^2,
\end{eqnarray}
where   $n_i$ is the
density of defects of type $i$.

  As in the usual pairbreaking theory,
\cite{AG,deg64,amb65,pa69},  $\hat{G}$ has the same form as does $\hat{G}_0$,
except that $\omega_n$ and $\Delta_0$  are replaced by their renormalized
equivalents $\tilde{\omega}_n$ and
$\tilde{\Delta}$, respectively.
We then obtain the {\it standard} equations for the renormalized gap and Matsubara frequency,
\begin{eqnarray}
\tilde{\omega}_n&=&\omega_n+(1/\tau_1+1/\tau_{\rm pb})
{{\tilde{\omega}_n}\over{2[\tilde{\omega}_n^2+\tilde{\Delta}^2]^{1/2}}},\label{omega}\\
\tilde{\Delta}&=&\Delta_0+(1/\tau_1-1/\tau_{\rm pb}){{\tilde{\Delta}}
\over{2[\tilde{\omega}_n^2+\tilde{\Delta}^2]^{1/2}}}\label{delta},\\
1/\tau_{\rm pb}&=&1/\tau_2+1/\tau_3\label{taupb}
\end{eqnarray}
is the total  pairbreaking rate. The new physics arises from $H_3$.
Evidently, within the self-consistent Born approximation,  the effects of
the random interactions are {\it exactly equivalent} to those of 
magnetic impurities.

Using standard pairbreaking theory, \cite{AG,deg64,amb65,pa69} one finds
\begin{eqnarray} 
{\omega_n\over {\Delta}_0}&=&u(1-{\zeta\over {\sqrt{1+u^2}}}),\label{dos}\\ 
u&=&\tilde\omega_n/\tilde\Delta,\\ 
\zeta&=&1/(\tau_{pb}\Delta_0). 
\end{eqnarray}
The spatial average gap $\Delta(T)$ is then 
\begin{eqnarray}
\Delta&=&|V|T\sum_{|\omega_n|\le\omega_0}\int{{d^2{\bf k}}\over{(2\pi)^3}}{\rm
Tr}[\rho_2\sigma_2\hat{G}({\bf k},\omega_n)],\nonumber\\ 
&=&\pi |V|N(0)T\sum_{|\omega_n|\le\omega_0} {1\over {\sqrt{1+u^2}}},\label{gap}
\end{eqnarray} 
leading
to the standard equation for $T_c/T_{c0}=t$,
\begin{eqnarray}
0&=&\ln(t)+\psi\Bigl({1\over{2}}+{{\alpha_{\rm pb}}\over{2\pi t}}\Bigr)
-\psi\Bigl({1\over{2}}\Bigr),\label{pb}\\
\alpha_{\rm pb}&=&1/(\tau_{\rm pb}T_{c0}),\label{alphapb}
\end{eqnarray}
where $\psi(x)$ is the digamma function.  For  small $\alpha_{\rm pb}$, $T_c\approx T_{c0} - 
\pi/4\tau_{\rm pb}$.  We note that  $T_c/T_{c0}$  can be suppressed
to zero   even {\it without any magnetic impurities}, for $1/\tau_3\ge1/\tau_{3c}=\pi T_{c0}/2\gamma$, where $\gamma=1.781$
is the exponential of Euler's constant.  In addition, the
superconductivity becomes gapless for
$1>\tau_{3c}/\tau_3\ge12e^{-\pi/4}\approx 0.912$.
 Thus,  even
an isotropic, $s$-wave superconductor can become gapless, as observed in
the cuprates with
STM.\cite{pan00,hudson01,HFK}

This can only occur in short coherence
length superconductors with strong local inhomogeneities in the pairing
interaction, as is a likely explanation for the vanishing of the $T_c$ in the
highly underdoped region of the cuprate phase diagram, although that region is
also complicated by the simultaneous appearance of local SDW/CDW order at the
non-superconducting regions not
included in this calculation.\cite{lang02}
 
In order to make direct comparison with STM experiments,
we  solve the T-matrix  for a single defect site. We assume the site has all three types of defects
associated with it.  We approximate the effects of the magnetic impurity by
assuming that its spin  behaves classically.\cite{shi68} 
Then, the T-matrix equation can be solved exactly,   
\begin{eqnarray}
\hat{T}(\omega_n)&=&{{\hat{V}(0)}\over{\hat{1}-\hat{g}_0(\omega_n)\hat{V}(0)}},\label{tmatrix}\\
\hat{V}(0)&=&\sum_{i=1}^3\hat{V}_i(0),\label{V}\\
\hat{g}_0(\omega_n)
&=&{{-\pi
N(0)}\over{[\omega_n^2+\Delta_0^2]^{1/2}}}[i\omega_n\rho_0\sigma_0+\Delta_0\rho_2\sigma_2],\label{g0}
\end{eqnarray} 
is the bare Green's function at the origin, $\hat{1}=\rho_0\sigma_0$ is the rank 4 identity matrix, and the effects
of a finite quasiparticle energy bandwidth have been neglected.
Bound states within the gap at $T=0$ at the frequency $\omega$ are 
obtained from
\begin{eqnarray}
{\rm det}|\hat{1}-\hat{g}_0(i\omega)\hat{V}(0)|&=&0\label{det}.
\end{eqnarray}

Solving Eqs. (\ref{V}) - (\ref{det})  exactly,
we generally obtain four bound states within
the gap at $\omega=\overline{\omega}_b\Delta_0$, where
\begin{eqnarray}
\overline{\omega}_b&=&\pm[A+sR]^{1/2}/B,\label{v1v2v3}\\
A&=&16v_2^2v_3^2+a_{+}^2(a_{-}^2+4v_1^2),\\
B&=&a_{+}^2+4v_2^2,\\
C&=&a_{+}^2-4v_3^2,\\
R&=&[A^2-B^2C^2]^{1/2},\\
a_{\pm}&=&v_1^2+v_3^2-v_2^2 \pm 1,\\
v_i&=& \pi N(0)U_i(0),
\end{eqnarray}
and $s=\pm1$.

It is useful to rewrite Eq. (\ref{v1v2v3}) for the three special cases of two
defects only.  For non-magnetic defects, $v_2=0$, there are only two  bound
states symmetric about zero bias,
\begin{eqnarray}
\overline{\omega}_b&\rightarrow&\pm{{\Bigl([(1+v_3)^2+v_1^2][(1-v_3)^2+v_1^2]\Bigr)^{1/2}}\over{(1+v_1^2+v_3^2)}}.\label{v1v3}
\end{eqnarray}

For the trivial case of a site with just a non-magnetic impurity, Eq. (\ref{v1v3}) shows
that there are no poles inside the gap.  However, for a pairing interaction
defect alone, there are poles at $\overline{\omega}_b=\pm|1-v_3^2|/(1+v_3^2)$.

For impurities alone, $v_3=0$, there are also only two bound states
symmetric about zero bias, at
\begin{eqnarray}
\overline{\omega}_b&\rightarrow&\pm{{|1+u_{+}u_{-}|}\over{[(1+u_{+}^2)(1+u_{-}^2)]^{1/2}}},\label{v1v2}
\end{eqnarray}
where $u_{\pm}=v_1\pm v_2$.
For $v_1=0$, this reduces to the result of Shiba for a classical magnetic
impurity, $\overline{\omega}_b=\pm|1-v_2^2|/(1+v_2^2)$, even though we did not
average over the classical spin direction before summing the T-matrix.\cite{shi68}

\begin{figure}[t]
\includegraphics[height=0.45\textwidth,angle=-90]{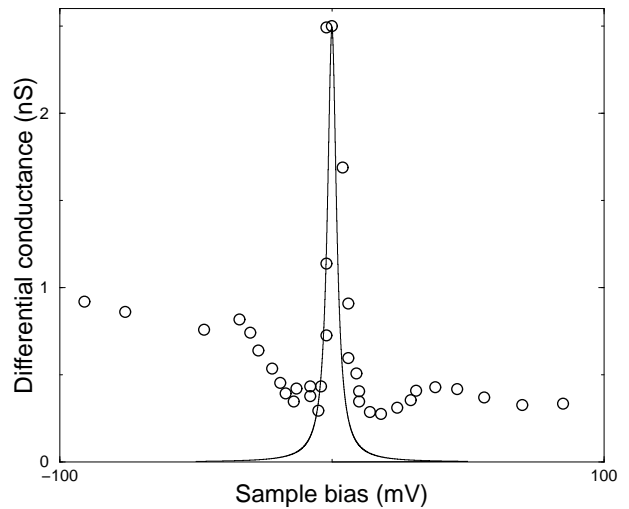}
\caption{Sketch of a bound state at zero bias with ${\bf v}=(0,0,\pm1)$, along with data  from
Ref. (9) for the STM differential conductance above a Zn site in BSCCO.} 
\end{figure}

The most interesting case arises when both $v_2,v_3\ne0$ and $v_3$  and
 $v_2\ne v_3$.  In the limit $v_1=0$, there are four bound states
 symmetric about zero bias at
 
\begin{eqnarray}
\overline{\omega}_b&\rightarrow&\pm{{(v_{+}^2-v_{-}^2+s|1-v_{+}^2v_{-}^2|)}
\over{(1+v_{+}^2)(1+v_{-}^2)}},
\label{v2v3}
\end{eqnarray}
where $s=\pm1$,  $v_{\pm}=v_2\pm v_3$. If $v_2, v_3\ne0$ and $v_2\ne v_3$, there 
are four  bound
states, exhibiting reflection symmetry about zero bias.  Otherwise, if either
$v_2$ or $v_3=0$, or if $v_2=v_3$, there are only two bound states that are
symmetric about zero bias. We note that Eq. (\ref{v2v3}) for either $v_2=0$ or
$v_3=0$ reduces to Eq. (\ref{v1v3}) and (\ref{v1v2}) with $v_1=0$, 
respectively.  In addition, for $v_2=v_3$, it reduces to Eq. (\ref{v1v2}) with
$v_1=0$ and $v_2\rightarrow 2v_2$, etc.  Moreover,
 setting
$v_2=v_3$ in Eq. (\ref{v1v2v3})  with $v_1$ arbitrary leads to only two bound
states symmetric about zero bias, at
$\overline{\omega}_b=\pm[|b_{-}|/b_{+}]^{1/2}$, where $b_{\pm}=(1+v_1^2)^2\pm4v_2^2$. Thus, we conclude that in order to
obtain four bound states, two on each side of zero bias, one requires $0\ne
v_2\ne 
v_3\ne0$.

When the defect is a quantum spin with a single component normal to the
surface, the spin operator $S_z$ commutes with the Hamiltonian, and the spin
states are easily described by $|SM\rangle$, with
$S_z|SM\rangle=M|SM\rangle$.  Then the magnetic impurity in the presence of
the non-magnetic potential and the pairing disorder can all be solved
exactly.  There are bound states for each of the $2S+1$ eigenstates.

In Figs. 1 and 2, we have illustrated how this solution can aid in 
understanding the STM results obtained  from the Bi sites
directly above 
Zn and Ni impurity sites  in the presumed top  underlying CuO$_2$ plane of
BSCCO, respectively.\cite{pan00,hudson01}  In Fig. 1, we fit the Zn STM data
with a single pole
obtained from Eq. (\ref{v1v2v3}) with ${\bf v}\equiv(v_1,v_2,v_3)=(0,0,\pm1)$ [or
equivalently with ${\bf v}=(0,\pm1,0)$] and a width
$\delta\omega$ chosen to fit the
 data.  The
 experimental  peak center appears at a  slight offset from zero, which 
can be understood quantitatively by adjusting $v_1<<1$ and a width that merges the
two resulting peaks into a single peak broader than the offest. 

\begin{figure}[t]
\includegraphics[height=0.45\textwidth,angle=-90]{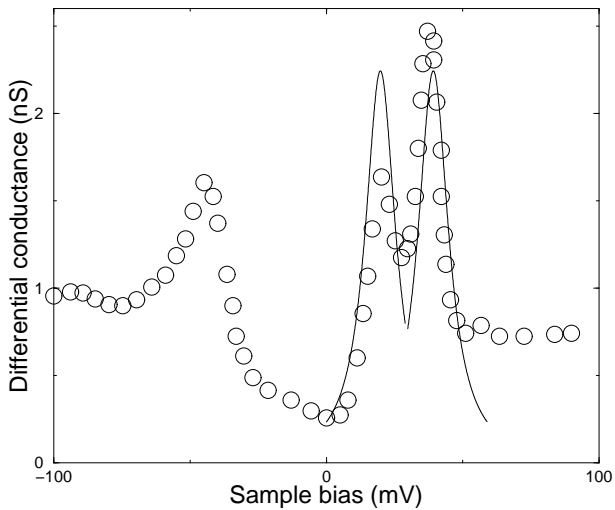}
\caption{Sketch of the two positive bias bounds states  obtained with
${\bf v}=(0, \pm0.51, \pm1.01)$, along with the data of Ref. (10) obtained  above
the Ni site in  BSCCO.} 
\end{figure}

 In Fig. 2,
we used Eq. (\ref{v2v3}) with ${\bf v}=(0, \pm0.51, \pm1.01)$ [or equivalently with
${\bf v}=(0, \pm1.01, \pm0.51)$] to fit the
two peak positions, and adjusted the widths to those of the data obtained
above a Ni site.\cite{hudson01}   We did not show the other two peaks that one expects from
Eq. (\ref{v2v3}).  That is because  the data not pictured here show that the two peaks
present in these data appear at equivalent negative biases on adjacent Bi
sites,\cite{hudson01} for reasons that are not particularly clear.  
In any event, our theory suggests that the STM data for Zn are consistent with
it behaving either as a strong pairing fluctuation defect or as a strong
magnetic impurity, and the Ni data suggests it
behaves  as both a strong magnetic impurity and a strong pairing defect 
in BSCCO.    The similar strengths of the
two pairbreaking defects might  explain why the $T_c$ suppressions
obtained in doping BSCCO with these elements are nearly identical.

Thus, we solved the T-matrix in this modified BCS model of
a local, on-site attractive pairing interaction with three types of defects on
a site.\cite{Nandini}  
For a superconductor with local, near-neighbor pairing of $d_{x^2-y^2}$-wave
symmetry, the local gap $\Delta_{ij}$ at the site $(i,j)$ on a tetragonal
lattice is coupled heirarchically to the $\Delta_{i'j'}$ at every
site $(i',j')$.  Hence, a generalization of our results to
$d$-wave superconductors would not be straightforward.

In summary, we have shown that disordered short coherence length
superconductors can exhibit pairbreaking  from spatial fluctuations in the pairing
interaction, in  a manner very similar to that found with magnetic
impurities. We studied the effects of a single site with up to three types of defects using the T-matrix
approximation, and found bound states within the superconducting gap arising
from either pairing fluctuations or magnetic impurities. Our
best fits to the scanning tunneling microscopy data above the sites of Zn and
Ni impurities suggest that Zn behaves  either as a strong pairing fluctuation defect
or as a strong magnetic impurity,
whereas Ni  behaves as both a strong magnetic impurity and a strong pairing defect.  


\end{document}